\documentstyle[11pt,newpasp,twoside,epsf]{article}
\begin{document}
\title{The Butcher-Oemler Effect in High Redshift X-ray Selected Clusters}
\author{David A. Wake, Chris A. Collins}
\affil{Astrophysics Research Institute, Liverpool John Moores University, Twelve Quays House, Egerton Wharf, Birkenhead CH41 1LD, UK.}
\author{Bruce W. Fairley, Laurence R. Jones}
\affil {School of Physics and Astronomy, University of Birmingham, Birmingham, B15 2TT, UK.}
\author{Doug J. Burke}
\affil {Harvard-Smithsonian Center for Astrophysics, 60 Garden Street, Cambridge, MA 02138, USA.}
\author{Robert C. Nichol, Kathy A. Romer}
\affil {Physics Department, Carnegie Mellon University, Pittsburgh, PA 15213, USA.}

\begin{abstract}
We are engaged in a wide-field, multi-colour imaging survey of X-ray selected clusters at intermediate and high redshift. We present blue fractions for the first eight out of 29 clusters, covering almost a factor of 100 in X-ray luminosity. We find no correlation of blue fraction with redshift or X-ray luminosity. The lack of a correlation with L$_{X}$, places strong constraints on the importance of ram-pressure stripping as a driver of the Butcher-Oemler effect.

\end{abstract}

\section{Introduction}

\begin{figure}
\plotfiddle{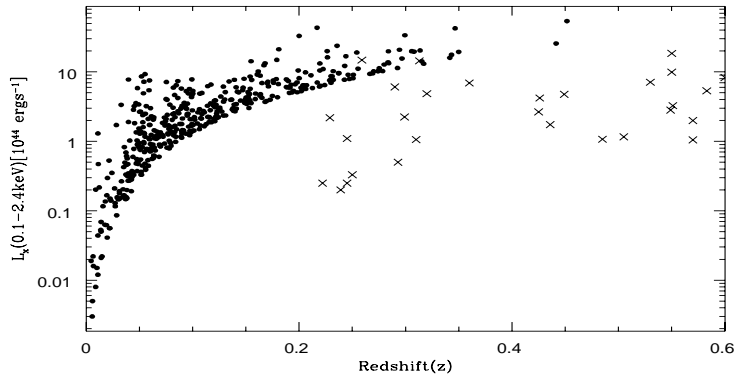}{1.5in}{0}{50}{40}{-150}{-25}
\caption{Redshift versus X-ray luminosity for our 29 clusters (crosses) and for the total REFLEX sample (filled circles).}
\end{figure}

Butcher \& Oemler (1984) found the fraction of blue galaxies $(f_{b})$, in their sample of centrally concentrated clusters, increased with redshift . The Butcher-Oemler (BO) effect was the first evidence for rapid galaxy evolution in clusters, and as such led to an extensive follow up program. Further photometric studies have generally confirmed the presence of this effect, although to varying degrees (e.g. Smail et al. 1998; Margoniner \& de Carvalho 2000; Kodama \& Bower 2001). Spectroscopic studies of the blue galaxies in question have revealed that they are cluster members, and have spectral features indicative of star-formation, often recently truncated (e.g. Dressler \& Gunn 1992). Morphological studies with HST indicate that these galaxies are disc dominated, with some evidence of enhanced interaction rates (Couch et al. 1994). These morphological studies also revealed that although the luminous elliptical galaxies present in local clusters remain in their high redshift counterparts, the large population of S0 galaxies are absent at high redshift. This has led to the association of these two effects with many mechanisms, such as ram-pressure stripping, harassment and tidal interaction,  being suggested for the transformation of the BO galaxies to S0s.
However a number of problems with these studies have been noted. In particular selection effects may be driving the observed BO effect (e.g. Andreon \& Ettori 1999). The clusters observed by Butcher and Oemler and by most of the additional BO studies have used optically selected cluster catalogues. These have tended to be selected in a single photometric band, so that as clusters are selected to higher redshift, bluer galaxies will become more important in the selection. This will bias clusters at high redshift to contain bluer galaxies. 
Another potential problem suffered by all studies to date, is caused by the tendency to select more and more rich/massive clusters with increasing redshift, leading to only the richest and most massive clusters being selected at high redshift . It is not yet clear whether $f_{b}$ depends on the cluster mass and how much this dependence may contribute to the BO effect.

\section{A Wide-Field Optical Imaging Survey of High Redshift X-ray Selected Clusters}
We are presently engaged in a broad-band photometric study of a well controlled sample of X-ray selected clusters, and present results on the blue fractions of the first eight observed here. Our clusters come from 4 catalogues, EMSS (Gioia et al. 1990), SHARC (Burke et al. 1997), BSHARC (Romer et al. 1999) and WARPS (Scharf et al. 1997). The flux limits of these surveys vary, with WARPS and SSHARC having very low flux limits allowing us to select clusters with low X-ray luminosities to high redshift. 
We have imaged a total of 29 clusters over two redshift intervals, $z\sim0.25$ and $z\sim0.5$, with a similar range in X-ray luminosities. Figure 1 shows a plot of the X-ray luminosities and redshifts of our cluster sample alongside those of the REFLEX survey (Bohringer et al. 2001). It is clear that our clusters have X-ray luminosities consistent with typical clusters in the local universe, allowing for easy and meaningful evolutionary comparisons to local samples.
We have imaged each of our clusters over a wide field in BVR for the $z\sim0.25$ sample, and VRI for the 
$z\sim0.5$ sample. These filters approximately map to rest-frame UBV for the chosen redshifts. The wide field allows an individual background to be calculated for each cluster taking into account line-of-sight large-scale structure variations. The imaging was done using the wide field mosaic cameras of the INT 2.5m, the CTIO 4m and the UH 2.2m, which have fields of view of 0.29, 0.1, and 0.36 deg$^2$ respectively.

\section{Results and Conclusions}

\begin{figure}
\begin{minipage}[b]{0.48\linewidth}
\plotone{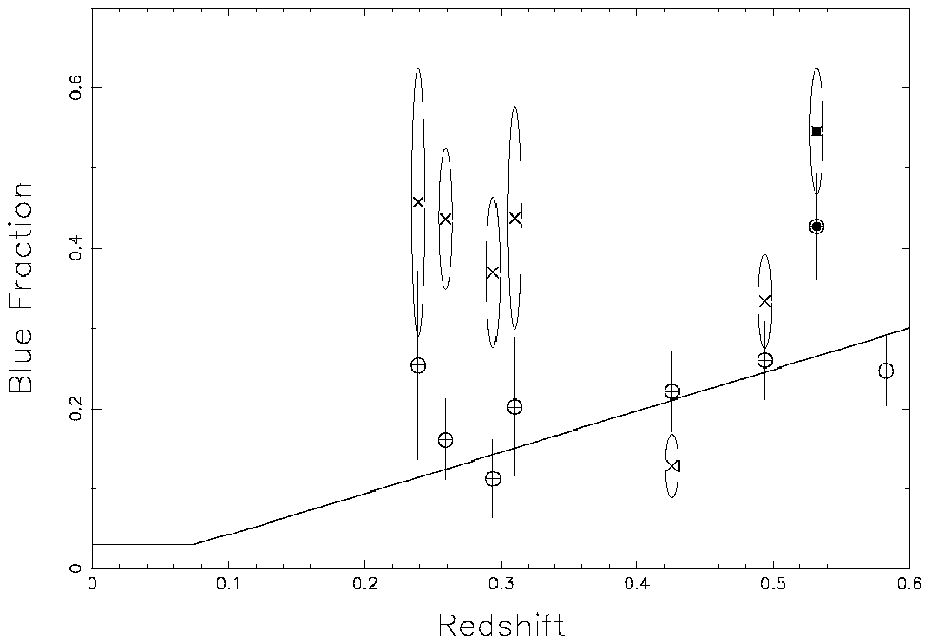}
\end{minipage}
\begin{minipage}[b]{0.48\linewidth}
\plotone{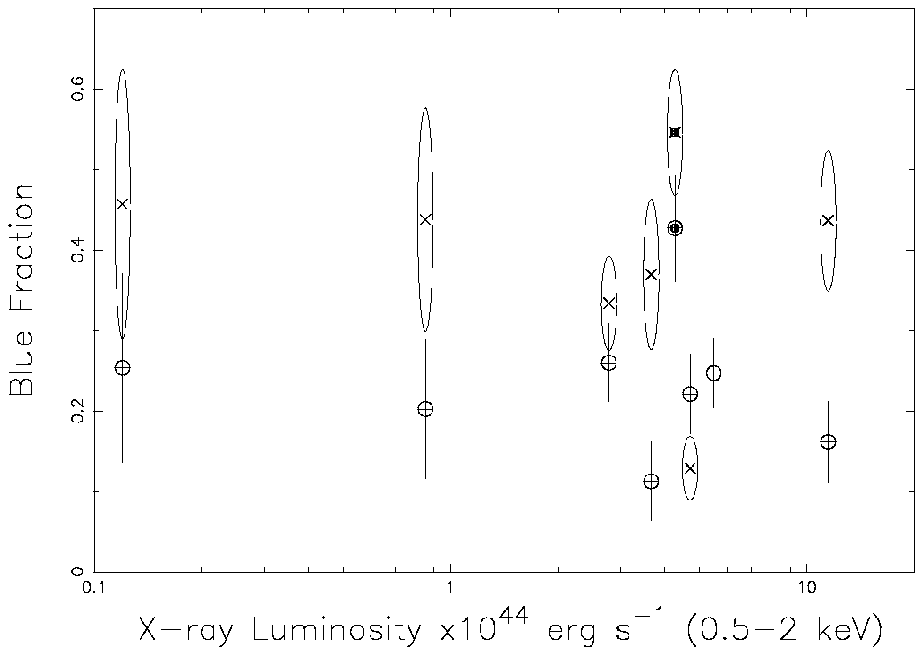}
\end{minipage}
\caption{Left plot shows blue fractions versus reshift and the right plot blue fraction versus X-ray luminosity for the eight clusters presented to $M_{V} = -20$, within $R_{30}$. See text for description of symbols.}
\end{figure}

Here we present blue fractions for 8 clusters imaged with the INT. We selected 4 in each redshift interval, with the lower redshift clusters chosen to have a wide range in L$_{X}$ and the high redshift sample very similar L$_{X}$. Our photometric accuracy is $\sim$0.03 mag, with completeness typically 24.2, 24.0, 22.7 mag for the three filters, with seeing between 0.79 and 1.52 arcsec.
Figure 2 shows both $f_{b}$ versus z and $f_{b}$ versus L$_{X}$, where $f_{b}$ is defined as the fraction of galaxies, within a characteristic radius $R_{30}$, 0.2 mag bluer than the early-type red-sequence, to a limiting magnitude $M_{V} = -20$ (the standard BO definition). The open circles shows the rest-frame \bv blue fraction, and the crosses show the rest frame \ub blue fraction. The solid circle and square are for RXJ2146.0+0423 which appears to be contaminated by a foreground group.
Immediately apparent from both graphs is the significant difference in the blue fractions measured in the two different colours, with the \ub blue fraction tending to be higher. This illustrates the importance of consistency when comparing blue fractions, and also illustrates the potential biases that may arise when selecting clusters in a single optical band.
No trend with either redshift or X-ray luminosity is apparent, although the small sample and large errors may be masking the trend. The original trend line of BO is plotted on the $f_{b}$-z graph and is not inconsistent with our data. The lack of a correlation with L$_{X}$ agrees with previous studies (e.g. Andreon \& Ettori 1999), and is shown to continue over a much wider range in L$_{X}$. This has particular consequences for the importance of ram-pressure stripping as a mechanism in the BO effect. The gas density in our poorest clusters, even in the core, is unlikely to provide effective stripping, whereas galaxies in our most X-ray luminous clusters should be rapidly stripped far from the core. As we see no dependence of blue fraction on L$_{X}$, this mechanism does not seem to be important in determining the blue fraction.
Figure 3 shows the blue fraction in rest frame \bv as a function of radius and absolute magnitude for our 8 clusters. We confirm the widely noted trend of increasing blue fraction with radius, although the rate of change varies greatly from cluster to cluster. For most, but not all, of our clusters we see an increasing blue fraction with fainter magnitude cut in agreement with Kodama \& Bower (2001). The radial trend of increasing $f_{b}$ is consistent with the idea of in-falling field galaxies being somehow transformed as they fall into the cluster. However, as previously mentioned it seems ram-pressure stripping is an unlikely mechanism for this transformation. Further details can be found in Fairley et al. (2001).

\begin{figure}
\plotfiddle{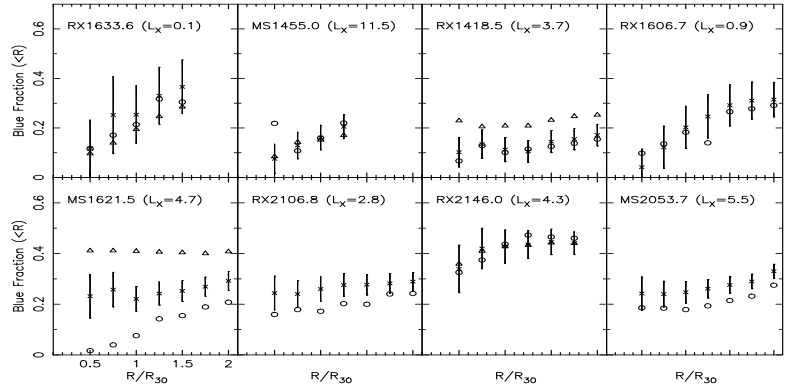}{1.6in}{270}{40}{28}{-150}{+140}
\caption{Blue fraction versus radius for all eight clusters in rest frame \bv. Open circles show blue fractions for $M_{V} < -21$, crosses for $M_{V} < -20$, and triangles for $M_{V} < -19$ }
\end{figure}

\end{document}